# CodeScore-R：用于评估代码合成功能准确性的自动化鲁棒指标

杨光[1]　周宇[1]　陈翔[2]　张翔宇[1]

[1](南京航空航天大学计算机科学与技术学院/人工智能学院　南京　211106)

[2](南通大学　信息科学技术学院,江苏　南通　226019)

[1](yang.guang@nuaa.edu.cn)

## CodeScore-R: An Automated Robustness Metric for Assessing the Functional Correctness of Code Synthesis

Yang Guang[1], Zhou Yu[1], Chen Xiang[2], and Zhang Xiangyu[1]

[1](*The College of Computer Science and Technology, Nanjing University of Aeronautics and Astronautics, Nanjing, China* 211106)

[2](*The School of Information Science and Technology, Nantong University, Nantong, China* 226019)

**Abstract** Evaluation metrics are crucial in the field of code synthesis. Commonly used code evaluation metrics can be classified into three types: match-based, semantic-based, and execution-based. Among them, the execution-based Pass@k metric accurately assesses the functionality of predicted code by executing test cases. However, calculating this metric requires a significant amount of overhead, necessitating the design of an automated evaluation metric that can assess the functionality of predicted code without the need for test cases. Additionally, a good evaluation metric should be robust, meaning it can maintain its accuracy even when the predicted code undergoes minor changes. To address these challenges, this paper proposes an automated robust metric called CodeScore-R, based on UniXcoder and contrastive learning, for evaluating the functionality of code synthesis. CodeScore-R employs techniques such as sketch-based processing, syntactic-equivalent transformations, and mutation testing to effectively mitigate the interference caused by identifiers, syntax structures, and operators on evaluation results. Experimental results demonstrate that in code generation and migration tasks in Java and Python, CodeScore-R outperforms other evaluation metrics and is more closely aligned with the Pass@k metric, while exhibiting stronger robustness.

摘要　评估指标在代码合成领域中至关重要．常用的代码评估指标可以分为三种类型：基于匹配，基于语义和基于执行．其中，基于执行的Pass@k指标通过执行测试用例，能够准确判断预测代码的功能准确性．然而，该指标的计算需要大量开销，因此亟需设计一种自动化评估指标，在无需测试用例时仍可评估预测代码的功能准确性．此外，好的评估指标应当具有鲁棒性，即预测代码发生微小改变时，评估指标仍能保持其准确性．为此，本文提出了一种基于UniXcoder和对比学习的自动化鲁棒指标CodeScore-R，用于评估代码合成的功能准确性．CodeScore-R采用草图化处理、语法等价转换和变异测试等技术手段，有效减轻了标识符、语法结构和运算符对评估结果的干扰．实验结果表明，在Java和Python语言上的代码生成和迁移任务中，CodeScore-R的表现优于其他无需测试用例的评估指标，且更接近Pass@k指标，并具有更强的鲁棒性．

关键词　代码合成评估指标；功能准确性；鲁棒性；代码合成；神经网络

中图法分类号　TP391

---

收稿日期：　；修回日期：

基金项目：国家自然科学基金面上项目(No. 61972197; No. 62372232)；江苏省研究生科研与实践创新计划项目(KYCX23_0396)

This work is supported by the National Natural Science Foundation of China (No. 61972197; No. 62372232), the Postgraduate Research & Practice Innovation Program of Jiangsu Province (KYCX23_0396).

通信作者：周宇(zhouyu@nuaa.edu.cn)

代码合成是一项自动化程序开发技术,旨在根据给定的规范和约束条件生成代码[1]。根据输入的模态,Ren 等人[2]将代码合成分为文本到代码 (text-to-code)和代码到代码 (code-to-code)两种类型[3]。其中,文本到代码是将自然语言描述转换为可执行代码的过程,例如代码生成[4];而代码到代码则是将现有代码转换为满足新需求的代码,例如代码迁移[5]。这两种技术都可以提高软件开发效率和质量,在实际应用中都具有重要的价值。此外,代码合成还可以为程序员提供一种更加友好的编程方式[6],并吸引那些在特定编程语言上经验有限的新手程序员参与软件开发。因此,代码合成对软件开发行业具有重要意义,并在其发展过程中变得越来越重要。

近年来,随着生成式 AI 模型[7]的快速迭代,代码合成领域取得了显著的发展,包括各种强大模型 (例如 CodeT5[8],CodeGen[9]和 CodeGeeX[10])的出现和多种数据集 (例如 Lyra[11],MBPP[12]和 CONCODE[13])的构建。尽管有这些令人振奋的进展,但代码合成的评价指标在现有的研究中受到的关注有限。评估指标在代码合成领域起着至关重要的作用,它们可以为研究人员和开发人员提供重要的指导和支持,帮助改进模型性能。目前,常用的代码评估指标可以分为三类:基于匹配、基于语义和基于执行的指标。基于匹配的指标,例如 BLEU[14]和 ChrF[15],将代码视为文本,并仅关注词素 (token)级别的匹配分数。基于语义的指标,例如 CodeBLEU[2],考虑到代码的句法结构和数据流信息作为代码的语义信息,并将其与词素信息混合计算。基于执行的指标,例如 Pass@k[16],通过人工编写测试用例,根据代码能否通过测试用例来评估模型的性能。其中,基于执行的 Pass@k 指标能更精准地判断预测代码的功能准确性,因此,本文以 pass@k 指标作为标准的功能准确性。但该指标在实际评估时,需要大量的人力、时间和成本开销,这在实际应用场景中并不总是可行的。因此,亟需设计一种自动化评估指标,在无需测试用例时仍可评估预测代码的功能准确性。

此外,一个好的评估指标应该具备鲁棒性,即当预测代码发生微小改变时,评估指标仍能保持准确性。然而现有的评估指标并不鲁棒,如图 1 所示,只有当预测代码 (b)与参考代码 (a)完全一致时,现有的指标才能给出正确的分数。然而,当代码中的用户标识符被修改或对代码进行了等价的语法转换时,即预测代码 (c)和代码 (d),此时代码 (c)、代码 (d)与参考代码 (a)的功能是一致的,但现有的指标无法识别它们,特别是 BLEU 指标,认为代码 (c)与参考代码 (a)的相似度仅为 0.040。另外,即使对代码进行微小的语义改变,即预测代码 (e),此时代码 (e)与参考代码 (a)的功能并不一致,但现有的评估指标仍无法正确识别它们,这种不鲁棒问题限制了现有评估指标的有效性。因此,本文对代码合成评估指标的鲁棒性检验提出了三种假设:

**假设 1**. 当生成的代码中发生用户自定义标识符替换时,指标分数应基本不变;

**假设 2**. 当生成的代码中发生语法等价转换时,指标分数应基本不变;

**假设 3**. 当对语义正确的代码进行语义变异时,原本得分高的指标应该大幅度降低.

这些假设旨在评估现有评估指标在应对代码改变时的鲁棒性,并为改进和开发更可靠的评估指标提供指导.

```
def sum ( a , b ):
    a = a + b
    return a                    参考代码 (a)

def sum ( a , b ):
    a = a + b
    return a                    预测代码 (b)

def f ( num_0 , num_1 ):
    num_0 = num_0 + num_1
    return num_0                预测代码 (c)

def sum ( a , b ):
    a += b
    return a                    预测代码 (d)

def sum ( a , b ):
    a = a - b
    return a                    预测代码 (e)
```

$BLEU$(a, b) = 1.0
$BLEU$(a, c) = 0.040
$BLEU$(a, d) = 0.653
$BLEU$(a, e) = 0.800

$ChrF$(a, b) = 1.0
$ChrF$(a, c) = 0.274
$ChrF$(a, d) = 0.807
$ChrF$(a, e) = 0.858

$CodeBLEU$(a, b) = 1.0
$CodeBLEU$(a, c) = 0.520
$CodeBLEU$(a, d) = 0.702
$CodeBLEU$(a, e) = 0.901

$CodeBERTScore$(a, b) = 1.0
$CodeBERTScore$(a, c) = 0.849
$CodeBERTScore$(a, d) = 0.969
$CodeBERTScore$(a, e) = 0.988

Fig. 1 Example of a robustness problem for code synthesis evaluation metrics

图 1 代码合成评估指标的鲁棒性问题示例图

近年来,评估代码合成的一个流行趋势是设计基于代码预训练语言模型[17] (pre-trained language models of code, CodePTMs). 相较于依赖词素层面的评估指标,通过提取代码内在的语义向量进行评估被认为是一个具有潜力的方向[18]. 本文提出了一种鲁棒的评估指标 CodeScore-R,它利用代码预训练语言模型 UniXcoder[19]来评估代码合成的功能准确性. 为了减轻用户标识符命名风格对指标的干扰,本文提出了一种基于抽象语法树 (abstract syntax tree, AST)的代码草图化处理方法,该方法从代码中提取用户自定义标识符,并用统一的占位符进行替换. 为了提升指标评估代码合成的功能准确性和鲁棒性,本文提出了针对代码的对比学习框架 ConCE (contrastive learning of code embedding). 该框架通过最小化正样本之间的语义距离和最大化负样本之间的语义距离来学习代码的表征. 通过语义向量之间的余弦相似度,可以评估代码合成的功能准确性. 在构造正样本方面,

CodeScore-R 通过进行语法等价转换来生成代码变体．而对于负样本，则通过利用变异测试中的运算符变异来对代码进行微小的改变．

为评估 CodeScore-R 的有效性，本文在代码生成与代码迁移任务上进行实验．在代码生成任务中，使用带有测试用例的 HumanEval[20]和 HumanEval-X[10]数据集，而在代码翻译任务中，本文使用 AVATAR[21]数据集，并通过众包的方式，手工构造了相应的测试用例．实验结果表明，对于 Java 和 Python 语言上的代码生成和迁移任务中，CodeScore-R 的表现优于现有评估指标，可以更接近 Pass@k 指标，并具有更强的鲁棒性．

本文的主要贡献如下：

1) 提出一种基于 UniXcoder 和对比学习的鲁棒指标 CodeScore-R，用于自动化评估代码合成功能准确性．
2) 引入草图化处理、语法等价转换和变异测试等技术，以减轻标识符、语法结构和运算符对评估结果的扰动．
3) 在 Java 和 Python 语言上的代码生成和迁移任务中，CodeScore-R 的表现优于现有评估指标，并具有更强的鲁棒性．

## 1 相关工作

当前已经有部分研究专注于不同的代码合成评估指标，这些指标对于衡量代码合成系统生成的代码质量至关重要．Liguori 等人[22]对现有代码合成评估指标和人工评估在生成攻击性代码 (offensive code) 的相似性进行了全面分析．他们的实验结果指出现有的指标通常不能提供准确的评估结果．目前，自动化代码评估指标主要可以分为三类：基于匹配、基于语义和基于执行的指标．此外，一些研究还通过人工评估的方法对预测代码进行分析．

### 1.1 基于匹配的评估指标

基于匹配的指标主要局限于对词素粒度的相似性计算，而忽略了代码的潜在语义信息．这类指标源自机器翻译、文本摘要等领域，包括 BLEU[14]，Rouge[23]和 ChrF[15]等指标．它们通过比较参考文本和预测文本之间的 n-gram 匹配程度来计算分数．此外，精准匹配度 (exact match，EM)指标也被广泛应用于代码合成任务[24]中．

然而，Eghbali 等人[25]指出，由于编程语言冗长的语法和编码约定，即使是完全不相关的代码片段也可能存在许多共同的 n-grams 信息．例如，在 Java 代码中，会出现大量的括号与分号的组合，这些冗余的信息可能导致 BLEU 指标的结果虚高．为了提高评估指标的准确性和可区分性，他们提出了 CrystalBLEU 指标，该指标在 BLEU 指标的基础上移除出现频率最高的 n-grams 信息．此外，Liguori 等人[22]认为，与其他基于匹配的指标相比，编辑距离 (edit distance，ED) 能够更好地衡量代码之间的相似性．

### 1.2 基于语义的评估指标

基于语义的指标考虑了代码的语法结构、数据流信息和潜在语义信息，但是仍不够鲁棒．Ren 等人[2]认为基于匹配的指标忽略了代码中的句法和语义特征，而 EM 指标过于严苛，无法识别出语义相同但语法不同的样本．为此，他们提出了 CodeBLEU 指标，它根据代码中的关键词进一步改进 BLEU 指标，并进一步通过 AST 注入代码语法，通过数据流注入代码语义．

Dong 等人[26]则基于 CodePTM 提出 CodeScore 指标，在带有测试用例的数据集上进行有监督学习，以此进行代码合成的功能性评估．然而，这种方法在训练时依赖于带有测试用例的数据集，成本较高．另一方面，Zhou 等人[27]利用 CodePTM 提取代码的潜在语义．具体而言，他们提出了 CodeBERTScore，利用 CodeBERT 模型[28]在大规模数据集上进行自监督学习，再对参考代码和预测代码进行上下文编码，计算每个词素之间的相似性分数．

### 1.3 基于执行的评估指标

基于执行的指标能准确判断预测代码的功能准确性，但需要投入大量的成本，包括测试用例的构造，解决运行环境的依赖问题，以及耗费大量时间进行测试用例的运行．

Kulal 等人[16]首次提出了基于测试用例的评估指标 pass@k，他们为每个问题生成了 K 个代码样本，并评估了 K 个样本中任意一个样本能够通过单元测试集的比率．另一方面，Liang 等人[11]则提出了 Code Executable 指标，用于判断预测代码是否能通过编译器的编译，该指标旨在没有测试用例的情况下观察预测代码的质量．

### 1.4 基于人工评估的方法

除了自动化评估指标外，人工评估也是一种可靠的方法，可以全面评估预测代码的准确性、语法、语义、复杂性、可读性、安全性等方面．然而，对每个编程问题进行手动评估是不切实际的，因为这需要大量的时间和成本．

因此，Yang 等人[29]提出了基于采样的方法，将该方法应用于小规模实证研究中．这种方法可以在可承受的时间和资源范围内获取对预测代码的高质量评估．

## 2 背景

### 2.1 问题定义

在代码合成任务中，通常面临以下问题：给定一个上下文 $x$(例如，自然语言功能描述或者源代码)，将 $x$ 输入到代码合成模型 $M$ 中，生成一个预测代码片段 $\hat{y}$，然后使用自动化评估指标来评估预测代码的质量．评估通常使用度量函数 $f(\hat{y}, y^*)$ 来比较预测代码 $\hat{y}$ 与参考代码 $y^*$．

$f(\hat{y}, y^*)$ 的值越大，表示预测代码与参考代码在功能准确性上越相似．在理想条件下，该度量函数应具备良好的鲁棒性，即满足前文提出的三个假设．具体来说，当预测代码 $\hat{y}_1$ 与参考代码 $y^*$ 之间语义一致，但词素不同 (记为 $\hat{y}_2$)或者语法上的实现不同 (记为 $\hat{y}_3$)时，评估指标计算的分数应该是一致的；当预测代码 $\hat{y}_4$ 与参考代码 $y^*$ 之间词素上相似但语义不一致时，评估指标应能正确识别出来差异．因此，本文希望提出一个满足 $f(\hat{y}_1, y^*) \approx f(\hat{y}_2, y^*) \approx f(\hat{y}_3, y^*) \gg f(\hat{y}_4, y^*)$ 的度量函数 $f$．

### 2.2 UniXcoder 介绍

UniXcoder[19]是一种基于 Transformer[30]的代码预训练语言模型，它在多项编程任务中表现出色．该模型使用 CodeSearchNet 数据集[31]进行了预训练，该数据集包含了来自六种编程语言 (Ruby，Java，Python，PHP，Go 和 Javascript)的约 2.3M 个代码样本．UniXcoder 通过使用三种不同类型的自注意力掩码策略来控制模型的行为，同时兼容编码、解码和编解码三种模式．除了基本的预训练任务，如掩码语言模型 (masked language modeling, MLM)，还提出了一些新的预训练任务，包括统一语言建模、去噪任务和代码片段表征学习等．

与其他 CodePTMs 相比，UniXcoder 能更全面地利用 AST 提供的代码结构信息，从而弥补了 AST 和词素表示的差异．此外，UniXcoder 还通过对比学习来学习代码片段的表征，使其在计算代码之间的相似度任务中更加适用．

## 3 CodeScore-R

本节主要介绍 CodeScore-R 指标的设计方法及具体实现，该指标的整体框架如图 2 所示．CodeScore-R 以 UniXcoder 为基座模型，并采用本文提出的 ConCE 框架进行自监督的代码表征学习．通过利用经过表征学习的 UniXcoder 模型，CodeScore-R 能够对生成的代码进行评估计算．接下来将介绍 CodeScore-R 的具体实现细节．

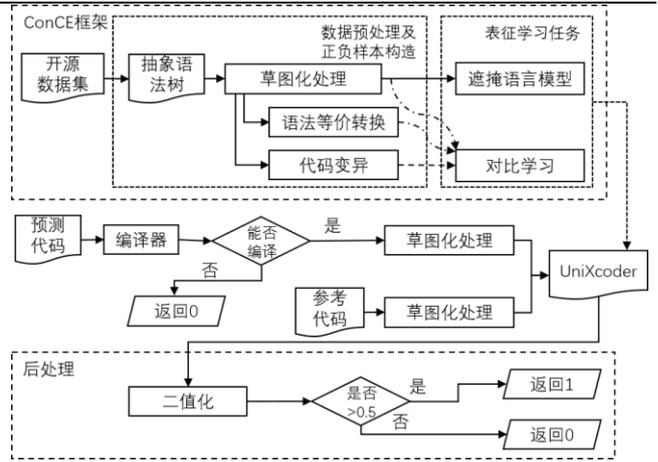

Fig. 2 CodeScore-R framework
图 2 CodeScore-R 框架图

### 3.1 ConCE 框架

ConCE 框架由数据预处理、正负样本构造以及表征学习任务三部分组成．

#### 3.1.1 数据预处理

为了降低不同标识符命名风格对评估指标的干扰，即为了满足假设 1，CodeScore-R 在数据预处理阶段对代码片段进行处理．首先，CodeScore-R 将代码片段解析为 AST，并根据预定义的规则识别出代码中用户自定义的标识符，例如函数名、参数名和变量名．然后，采用草图化的方法统一变量命名风格．图 3 展示了预处理前后的对比示例图．由于参数、变量及函数名的命名与程序员的编码习惯相关，这可能会对评估指标的计算产生干扰．但是，对参数、变量及函数名进行草图化处理并不会影响代码的语法和语义信息．

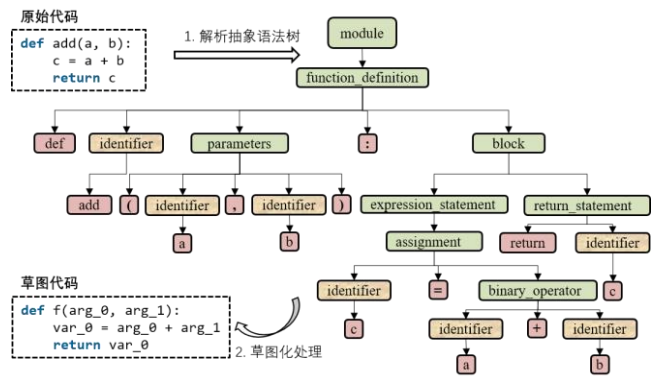

Fig. 3 An Example of Code Pre-processing
图 3 代码预处理示例图

具体来说，CodeScore-R 利用 tree-sitter 工具[1]对代码片段进行解析，并根据每个类型为 identifier 的叶子节点的父节点进行判断，以识别出函数名、参数名和变量名的词素信息．草图化是将这些词素以统一的编码方式进行逐一替换．例如，函数名替换为"f"，

---

[1] https://github.com/tree-sitter

参数名和变量名分别按照先后顺序替换为{arg_0, arg_1, …, arg_m}和{var_0, var_1, …, var_n}. 在后续的正负样本构造及表征学习中，CodeScore-R 都是使用草图代码 (sketch code)而不是原始代码 (origin code)进行实验，以尽可能地避免标识符词素对评估指标的影响.

#### 3.1.2 正负样本的构造

为了进行后续的表征学习，CodeScore-R 需要构造正负样本. 对于正样本的构造，CodeScore-R 借鉴了 SimCSE[32]中的思想，一方面采用两次不同的 dropout 方法生成正样本，以增强其对代码语义的理解和鲁棒性. 另一方面，本文根据代码的语法特性，对代码进行语法上等价的转换，生成与原始代码功能相同但形式不同的代码片段. 这种转换可以包括改变代码中的控制流结构、使用不同的语法特性等. 通过这样的等价转换，可以扩展训练数据集，提供更多样化的正样本供 CodeScore-R 进行学习. 为了确保生成的语法等价的代码的质量，选择了代码重构[33]中常见的四种规则进行操作，具体规则见表 1.

Table 1 Relus of Code Syntax Equivalence Transformation

表 1 代码语法等价转换规则表

| 规则 | 代码示例 |
| --- | --- |
| 循环交换(loop exchange) | For ⇔ While |
| 表达式交换(expression exchange) | a+=b ⇔ a=a+b |
| 判断体交换(permute exchange) | if(a){A}else{B} ⇔ if(!a){B}else{A} |
| 判断条件交换(condition exchange) | a>b ⇔ b>a or True ⇔ !False |

循环交换是将 for 循环替换为等效的 while 循环或将 while 循环替换为等效的 for 循环. 这样的交换可以改变循环结构的表达方式，但保持了代码功能的等效性. 表达式交换是基于运算符特性进行等价替换，例如将 "a+=b" 替换为 "a=a+b". 这种交换在语义上保持了表达式的等效性，但改变了表达式的形式. 判断体交换和判断条件交换是常见的代码重构规则，用于条件语句的等价转换. 判断体交换将条件语句中的 if 分支和 else 分支进行交换，而判断条件交换交换条件表达式中的两个子表达式. 这样的交换保持了条件语句的等效性，但改变了条件语句的结构或条件的顺序.

对于负样本的构造，CodeScore-R 采用了变异测试的思想. 变异测试[34]是一种软件测试方法，通过对原始代码进行变异 (如改变运算符、改变变量引用等)，生成具有细微差异的代码，从而检验测试用例的充分性和代码的健壮性. 在构造负样本时，CodeScore-R 也使用了类似的方法. 它通过对原始代码进行运算符的变异操作，例如将加法操作变异为减法操作或将逻辑运算符变异为移位运算符等，可以生成与原始代码形式上相似但在语义上存在异的负样本，具体变异规则见表 2. 通过引入这样的负样本，CodeScore-R 能够区分细微差异的代码，并提升对代码变异的鲁棒性.

Table 2 Rules of Code Mutation Testing

表 2 变异测试规则表

| 规则 | 运算符 |
| --- | --- |
| 算术运算符 | +, -, *, /, **, % |
| 关系运算符 | >, <, >=, <=, ==, != |
| 条件运算符 | &&, \|\|, &, \|, ^ |
| 移位运算符 | <<, >>, >>> |
| 逻辑运算符 | &, \|, ^ |
| 赋值运算符 | =, +=, -=, *=, /=, %=, **=, <<=, >>=, >>>= |

#### 3.1.3 表征学习任务

为了进行表征学习，CodeScore-R 采用了两个自监督预训练任务，分别是掩码语言模型和对比学习，如图 4 所示. 这些任务旨在让模型学习代码的语义信息，并能够区分代码之间的微小差异.

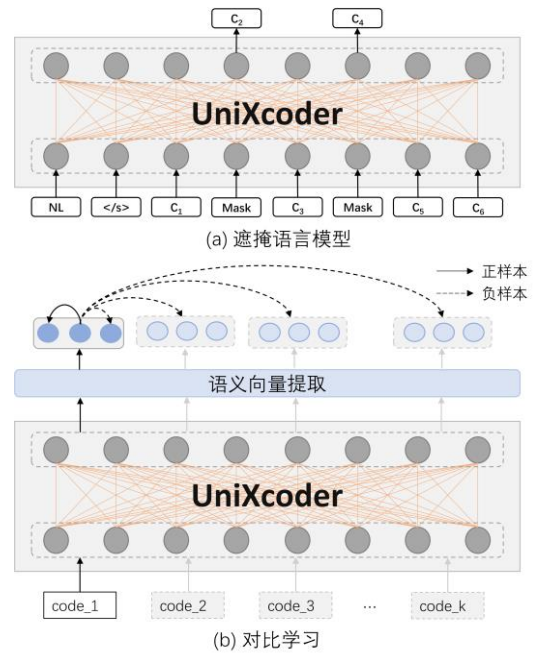

Fig. 4 Code Representation Learning Tasks

图 4 表征学习任务示例图

1)掩码语言模型. 掩码语言模型的任务是最大化正确预测被屏蔽词素的概率，以使 CodeScore-R 能够学习到代码的语义表示. 在数据预处理阶段，由于对代码进行了草图化处理，导致草图代码的词素失去了原始代码中部分语义信息. 为了确保模型能够学习到完整的语义信息，本文引入了代码对应的注释 NL，并随机选择草图代码中 15% 的词素进行掩码操作. 具体而言，该任务以一定的概率对选定的词素进行替

换，其中，80%的概率将其直接替换为特殊标记"<mask>"，10%的概率将其随机替换为其他词素，而剩下的 10%则保持不变. 通过这种方式，模型需要预测被掩码的词素，从而学习到代码的语义信息.

给定一个 NL-Code 对的数据 ($d =\{w, c\}$)作为输入，其中 $w$ 是 NL 的序列，$c$ 是代码词素的序列，可以将掩码操作定义为：

$$m_i^c \sim \text{unif}\{1, |c|\} \text{ for } i = 1 \text{ to } |c| \quad (1)$$

掩码语言模型的目标是预测被屏蔽掉的原始标记，可以定义为：

$$\mathcal{L}_{MLM}(\theta) = \sum_{i \in m_i^c} -\log p(d_i | c^{\text{masked}}) \quad (2)$$

2)对比学习. 对比学习任务旨在使模型能够区分正样本和负样本，以增强对微小差异的敏感性. 对于正样本的构造，本文采用 3.1.2 节提出的两种方法：50%的概率通过两次 dropout 得到正样本，另外 50%的概率通过代码语法等价转换得到正样本. 这样的设计既避免了模型对输入长度的敏感性问题，又增加了模型对不同语法的识别能力，提高了模型的语法鲁棒性. 对于负样本的构造，同样采用 3.1.2 节提出的基于变异测试的方法. 考虑到正负样本的构造都是基于多种规则的，为了提升样本的多样性，本文使用组合的方式将这些规则进行融合，再分别随机抽选一个样本当作正样本和负样本.

给定一个代码片段 $c_i$，我们将 $c_i^+$ 记为正样本，将 $c_i^-$ 记为负样本，一起组合为 ($c_i$，$c_i^+$，$c_i^-$)，利用 UniXcoder 对它们进行语义向量提取，得到向量 ($h_i$，$h_i^+$，$h_i^-$). 具体的提取方法是先对输入的代码进行上下文编码得到其隐藏向量 $\boldsymbol{H} \in \mathbb{R}^{batch \times N \times d}$，其中 $N$ 表示为输入代码的长度，$d$ 表示为向量的维度. 接着提取第一个词素，即[CLS]的向量值，并经过一层 ReLU 层进行激活，最终作为代码的语义表征 $\boldsymbol{h}$. 对比学习任务的损失函数定义为：

$$\mathcal{L}_{CL}(\theta) = -\log \frac{e^{\text{sim}(\boldsymbol{h}_i, \boldsymbol{h}_i^+)/\tau}}{\sum_{j=1}^{batch}\left(e^{\text{sim}(\boldsymbol{h}_i, \boldsymbol{h}_j^+)/\tau} + e^{\text{sim}(\boldsymbol{h}_i, \boldsymbol{h}_j^-)/\tau}\right)} \quad (3)$$

其中 batch 代表着批次的大小，sim 代表着余弦相似度. 最终，ConCE 框架的优化目标可设置如下：

$$\mathcal{L}(\theta) = \mathcal{L}_{MLM}(\theta) + \mathcal{L}_{CL}(\theta) \quad (4)$$

### 3.2 评估流程

CodeScore-R 的评估流程如下：首先，以预测代码和参考代码作为输入. 为了避免预测代码包含语法错误，CodeScore-R 首先借助编译器判断预测代码是否能够编译通过. 如果预测代码无法通过编译，则直接输出功能准确性分数为 0. 如果预测代码通过了编译，则对预测代码和参考代码进行草图化处理，以减轻标识符对评估结果的影响.

接着，CodeScore-R 借助经过表征学习后的 UniXcoder，分别提取预测代码和参考代码的语义表征向量，并进行余弦相似度计算. 通常来说，余弦相似度计算结果的值域为[-1, 1]. 然而，由于在语义表征向量提取过程中使用了 ReLU 激活函数，它保持非负值不变. 因此，经过 ReLU 激活后的语义表征向量中的每个元素都是非负的.

余弦相似度的计算公式如下：

$$\text{cosine\_similarity}(x, y) = \frac{x \cdot y}{||x|| \cdot ||y||} \quad (5)$$

其中，x·y 表示向量 x 和 y 的内积，||x||和||y||分别表示向量 x 和 y 的范数. 由于语义表征向量中的元素都是非负的，内积和范数的结果也是非负的. 因此，余弦相似度的分子是非负的内积，分母是非负的范数乘积，所以余弦相似度的结果也是非负的，即值域为[0, 1].

为了更好地拟合 pass@k 指标，CodeScore-R 加入了后处理操作，即对相似度分数进行二值化处理，最终输出预测代码的功能准确性分数：若相似度分数大于 0.5，输出为 1；否则输出为 0.

## 4 实验设置

为了验证论文提出的 CodeScore-R 的有效性，本文设计了如下五个实验问题.

RQ1：与现有指标相比，CodeScore-R 指标对功能准确性的拟合程度如何？

RQ2：与现有指标相比，CodeScore-R 针对代码标识符扰动下的鲁棒性如何？

RQ3：与现有指标相比，CodeScore-R 针对代码语法扰动下的鲁棒性如何？

RQ4：与现有指标相比，CodeScore-R 针对代码语义扰动下的鲁棒性如何？

RQ5：不同的语义提取方法对 CodeScore-R 有什么影响？

### 4.1 数据集

本文以 CodeSearchNet 数据集[31]作为表征学习的数据集，具体来说，我们提取其中的 Java 代码和 Python 代码以及它们相应的功能注释，并对它们进行预处理和正负样本的构造. 由于不是每个代码都能根

据规则转换为相应的变体代码，因此过滤出那些无法构造正负样本的代码以作为最终的预训练数据集.

对于下游任务，本文在两种编程语言 (Java 和 Python)任务上评估了 CodeScore-R，包括代码生成任务和代码迁移任务.

代码生成任务是根据用户的功能描述加上代码签名 (signature)来生成相应的代码. 本文使用由 Python 语言构造的 HumanEval 数据集[20]和包含了 Java 语言的 HumanEval-X 数据集[10]作为实验对象，它们由包含了 164 条带有测试用例的样本构成的.

代码迁移任务是根据一种编程语言的源代码生成另一种编程语言的目标代码. 本文使用 AVATAR 数据集[21]，其中包含了挖掘自在线编程网站上的 Java 和 Python 的代码对. 我们从中随机挑选了 200 对数据并手工构造测试用例，为了保证测试用例的质量，我们通过众包的方式聘请具有 2-3 年开发经验的工程师，并为每个代码段构造 5 条测试用例以确保测试用例的多样性.

### 4.2 基准指标

本文选择 10 种代码合成评估的基准指标，包括 BLEU[14]，ROUGE-L[23]，ChrF[15]，ED[22]，WeightBLEU[2]，SyntaxMatch[2]，DataflowMatch[2]，CodeBLEU[2]，CrystalBLEU[25] 和 CodeBERTScore[27]. 这些指标被广泛使用在代码合成的相关研究中. 其中，BLEU，ROUGE-L，ChrF，ED 和 CrystalBLEU 是基于匹配的评估指标. CodeBLEU 和 CodeBERTScore 是基于语义的评估指标，WeightBLEU，SyntaxMatch，DataflowMatch 是 CodeBLEU 计算过程中设计出的指标，分别对应了改进的 BLEU，语法匹配与数据流匹配的计算.

### 4.3 评估设置

为了评估现有评估指标和代码功能准确性之间的拟合程度，现有的研究通常选择相关系数指标[35] (如 Pearson 相关系数、Spearman 相关系数和 Kendall 相关系数)和拟合指标[36] (如平均绝对误差). 其中，相关系数是用来衡量两个变量之间线性相关程度的指标，而拟合指标则是用来衡量模型的拟合程度的指标.

本文的目标是检验 CodeScore-R 能否自动评估预测代码的功能准确性，即是否能拟合 pass@1 指标的结果，因此本文选择平均绝对误差 (mean absolute error，MAE)评估它们之间的绝对误差. MAE 值越大，说明拟合程度越差；反之则说明拟合程度越好. MAE 可以定义为：

$$\text{MAE} = \frac{\sum_{i=1}^{N} |M_i^1 - M_i^2|}{N} \quad (6)$$

### 4.4 实现细节

在后续的实证研究中，所有实验均运行在 Windows 10 操作系统上，其内存为 32GB，CPU 处理器为 Intel Core i7-9750H. 对于编译器，本文使用 Python 3.9.1 版本[2]作为 Python 代码的编译器，并使用 JDK 18.0.2.1 版本[3]编译 Java 代码.

随着语言模型和训练数据的规模不断增长，基于大型语言模型的生成式 AI 表现出各种涌现行为. 其中一种能力是零样本学习，它允许模型在特定指令或提示中回答. 这些生成式 AI 模型取得了出色的性能，并在代码合成任务上也展示了广阔的潜力. 因此，本文通过调用 ChatGPT 的接口[4] (gpt-3.5-turbo)完成代码生成与代码迁移任务. 为了保证结果的稳定性，本文将接口中的 temperature 参数设置为 0，并只返回 1 个代码结果，用于计算 pass@1 指标.

经测试，ChatGPT 在 Java 代码生成任务上的 pass@1 指标为 68.29%，在 Python 代码生成任务上的 pass@1 指标为 71.34%；ChatGPT 在 Python-to-Java 代码合成任务上的 pass@1 指标为 80.5%，在 Java-to-Python 代码合成任务上的 pass@1 指标为 86.5%.

## 5 结果分析

### 5.1 RQ1 结果分析

RQ1：与现有指标相比，CodeScore-R 指标对功能准确性的拟合程度如何？

为了评估 CodeScore-R 指标的拟合程度，本文选用了机器翻译领域以及代码合成领域中常用的指标作为基准评估指标. 表 3 给出了 CodeScore-R 指标和选择的基准方法与功能准确性之间的拟合结果，我们对最佳结果进行了加粗标记.

结果显示，与所有基准指标相比，CodeScore-R 指标在所选的四个下游任务数据集上对功能准确性的拟合更好. 在 Java 代码生成、Python 代码生成、Python-to-Java 代码迁移和 Java-to-Python 代码迁移任务中，与最优的基准指标相比，CodeScore-R 指标的拟合误差分别降低了 7.31%、11.42%、13.37% 和 17.22%. 其中，在代码迁移任务上的拟合程度普遍优

---

[2] https://www.python.org/downloads/release/python-391/
[3] https://www.oracle.com/java/technologies/javase/18-0-1-relnotes.html
[4] https://platform.openai.com/docs/models/gpt-3-5

于在代码生成任务上的拟合程度；在 Python 编程语言上的拟合程度普遍优于在 Java 编程语言上的拟合程度.

此外，表 3 的结果指出，基于 LLM 的评估指标普遍优于其他的基准指标，这也体现出该类指标的优势和可行性. 在基于匹配的基准指标中，ED 的拟合结果最优，这与之前的研究结论[22]一致，而在基于语义的基准指标中，CodeBERTScore 则表现出最好的拟合结果.

Table 3 Results between CodeScore-R and Baseline Metrics

表 3 CodeScore-R 指标与基准指标的 MAE 对比结果

| 指标 | 代码生成 | | 代码迁移 | |
|---|---|---|---|---|
| | Java | Python | Java | Python |
| BLEU | 0.453 | 0.498 | 0.268 | 0.347 |
| ROUGE-L | 0.391 | 0.409 | 0.225 | 0.192 |
| ChrF | 0.395 | 0.424 | 0.229 | 0.262 |
| ED | 0.392 | 0.401 | 0.208 | 0.198 |
| WeightBLEU | 0.415 | 0.467 | 0.252 | 0.329 |
| SyntaxMatch | 0.404 | 0.436 | 0.283 | 0.420 |
| DataflowMatch | 0.398 | 0.415 | 0.217 | 0.313 |
| CodeBLEU | 0.418 | 0.454 | 0.255 | 0.352 |
| CrystalBLEU | 0.401 | 0.432 | 0.235 | 0.220 |
| CodeBERTScore | 0.342 | 0.324 | 0.202 | 0.151 |
| CodeScore-R | **0.317** | **0.287** | **0.175** | **0.125** |

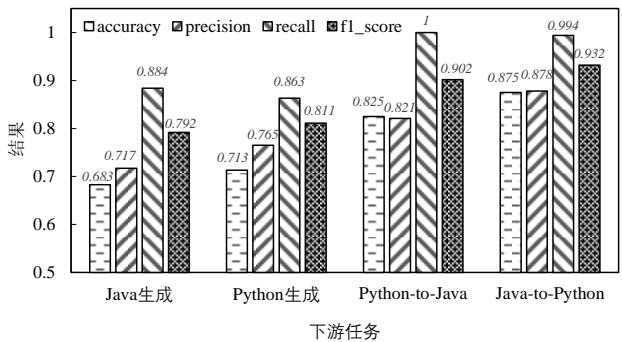

Fig. 5 Classification Results of CodeScore-R and Functional Correctness

图 5 CodeScore-R 与功能准确性的分类结果

考虑到 CodeScore-R 指标输出的是 0-1 值，因此相较于其他指标，我们可以将 CodeScore-R 的输出与功能准确性视作分类任务，通过比较 CodeScore-R 的输出与功能准确性之间的准确率、精度、召回率和 F1-Score 以进一步评估 CodeScore-R 指标的性能. 同时可以分析表 3 结果中的 MAE 指标是否存在假阳性或者假阴性数据过多而导致的误差虚低的情况. 结果如图 5 所示，可以看到，在代码生成任务上，CodeScore-R 的 F1-Score 值在 0.8 左右，而在代码迁移任务上，CodeScore-R 的 F1-Score 值能在 0.9 以上，该结果可以进一步表明 CodeScore-R 指标的准确性.

## 5.2 RQ2 结果分析

RQ2：与现有指标相比，CodeScore-R 针对代码标识符扰动下的鲁棒性如何？

基于 RQ1 中的分析结果，我们发现 CodeScore-R 指标在所有任务中对功能准确性的拟合程度超过了本论文考虑的十种基准指标. 然而，一个好的评估指标应具备鲁棒性. 为了研究 CodeScore-R 指标与现有基准指标在代码标识符扰动下的鲁棒性（即是否满足假设 1），以及我们提出的代码草图化处理的有效性，我们进行了两组对比实验：针对 Origin-to-Sketch (O2S) 的对比实验和针对 Sketch-to-Sketch (S2S) 的对比实验.

Origin-to-Sketch (O2S) 是仅对预测代码进行草图化处理，而参考代码保持不变. 由于草图化处理会识别代码中的所有用户自定义标识符并进行统一替换，我们将其视为代码标识符扰动的一种极端情况.

Sketch-to-Sketch (S2S) 则是对参考代码和预测代码都进行草图化处理，以探究草图化处理是否能缓解对指标的影响.

Table 4 Comparison Results of CodeScore-R with Baselines for Token Perturbation on Code Generation Tasks

表 4 CodeScore-R 与基准指标在代码生成任务上针对词素扰动的对比结果

| 指标 | Java 代码生成 | | Python 代码生成 | |
|---|---|---|---|---|
| | O2S | S2S | O2S | S2S |
| BLEU | 0.544 | 0.435 | 0.593 | 0.465 |
| ROUGE-L | 0.501 | 0.390 | 0.444 | 0.398 |
| ChrF | 0.483 | 0.376 | 0.537 | 0.398 |
| ED | 0.429 | 0.381 | 0.454 | 0.385 |
| WeightBLEU | 0.522 | 0.394 | 0.574 | 0.432 |
| SyntaxMatch | 0.404 | 0.401 | 0.450 | 0.439 |
| DataflowMatch | 0.402 | 0.395 | 0.422 | 0.409 |
| CodeBLEU | 0.468 | 0.406 | 0.510 | 0.436 |
| CrystalBLEU | 0.486 | 0.389 | 0.508 | 0.405 |
| CodeBERTScore | 0.376 | 0.340 | 0.371 | 0.316 |
| CodeScore-R | - | **0.317** | - | **0.287** |

Table 5 Comparison Results of CodeScore-R with Baselines for Token Perturbation on Code Migration Tasks

表 5 CodeScore-R 与基准指标在代码迁移任务上针对词素扰动的对比结果

| 指标 | Python-to-Java | | Java-to-Python | |
|---|---|---|---|---|
| | O2S | S2S | O2S | S2S |
| BLEU | 0.590 | 0.265 | 0.718 | 0.340 |

| | | | | | | | | | |
|---|---|---|---|---|---|---|---|---|---|
| ROUGE-L | 0.485 | 0.227 | 0.310 | 0.191 | ChrF | 0.385 | 0.421 | 0.251 | 0.318 |
| ChrF | 0.493 | 0.223 | 0.598 | 0.242 | ED | 0.391 | 0.393 | 0.238 | 0.230 |
| ED | 0.335 | 0.208 | 0.380 | 0.198 | WeightBLEU | 0.412 | 0.444 | 0.300 | 0.362 |
| WeightBLEU | 0.579 | 0.250 | 0.709 | 0.323 | SyntaxMatch | 0.414 | 0.448 | 0.304 | 0.444 |
| SyntaxMatch | 0.283 | 0.283 | 0.420 | 0.411 | DataflowMatch | 0.396 | 0.414 | 0.221 | 0.324 |
| DataflowMatch | 0.241 | 0.218 | 0.350 | 0.304 | CodeBLEU | 0.418 | 0.446 | 0.286 | 0.379 |
| CodeBLEU | 0.423 | 0.254 | 0.549 | 0.345 | CrystalBLEU | 0.430 | 0.447 | 0.316 | 0.369 |
| CrystalBLEU | 0.474 | 0.213 | 0.449 | 0.211 | CodeBERTScore | 0.347 | 0.320 | 0.219 | 0.167 |
| CodeBERTScore | 0.282 | 0.201 | 0.273 | 0.149 | CodeScore-R | **0.311** | **0.278** | **0.182** | **0.129** |
| CodeScore-R | - | **0.175** | - | **0.125** | | | | | |

表 4 和表 5 分别展示了 CodeScore-R 指标和所选基准方法在代码生成和代码迁移任务上的评估结果，我们对最佳结果进行了加粗标记．从结果中我们可以观察到，在 RQ2 的分析结果中，除了 CodeScore-R 指标之外，现有基准指标对于代码词素扰动普遍不具备鲁棒性．以 Java-to-Python 代码迁移任务为例，在原始设置下，其 BLEU 指标的 MAE 误差值为 0.347．然而，在 O2S 设置下，即仅对预测代码进行草图化处理的情况下，该值增加至 0.718，即拟合误差增加了 106.92%．而在 S2S 设置下，即对参考代码和预测代码都进行草图化处理的情况下，该值降至 0.340，与原始设置相比，误差减少了 2.02%．这表明草图化处理能够提升现有基准指标的评估准确性和鲁棒性．在后续的 RQ3 和 RQ4 中，为了公平比较 CodeScore-R 和现有基准方法的鲁棒性，我们提前对参考代码和预测代码都进行了草图化处理．

### 5.3 RQ3 结果分析

RQ3：与现有指标相比，CodeScore-R 针对代码语法扰动下的鲁棒性如何？

为了探究 CodeScore-R 指标与现有基准指标对代码语法扰动下的鲁棒性，即是否满足假设 2，我们根据表 1 中提出的 4 种规则，并通过相关组合构造出多个语法转换方法．在此 RQ 中，为了避免实验的偶然性，我们选择了五个不同的随机种子，并为每个预测代码生成了五种变体代码．然后，我们计算每个变体代码与参考代码之间的评估指标，并在表 6 中给出了这些指标的均值．我们对最佳结果进行了加粗标记．

Table 6 Comparison Results of CodeScore-R with Baselines for Syntax Perturbation

表 6 CodeScore-R 指标与基准指标针对 syntax 扰动的对比结果

| 指标 | 代码生成 | | 代码迁移 | |
|---|---|---|---|---|
| | Java | Python | Java | Python |
| BLEU | 0.451 | 0.479 | 0.318 | 0.388 |
| ROUGE-L | 0.453 | 0.493 | 0.377 | 0.502 |

从表 6 的结果中可以观察到，相对于代码中的词素扰动，现有指标在面对语法扰动时受到的影响较小．然而，ROUGE-L 指标受语法扰动的影响最为显著，在 Java-to-Python 代码迁移任务中，其 MAE 值从 0.192 增加到 0.502，即误差增加了 161.46%．与此相比，本文提出的 CodeScore-R 对语法扰动的影响较小，并且在语法扰动下，其 MAE 拟合结果仍然是最佳的．这些表 6 的结果进一步证明了本文提出的 ConCE 中表征学习方法的有效性．

### 5.4 RQ4 结果分析

RQ4：与现有指标相比，CodeScore-R 针对代码语义扰动下的鲁棒性如何？

为了探究 CodeScore-R 指标与现有基准指标对代码语义扰动下的鲁棒性，即是否满足假设 3，我们根据表 2 中提出的 6 种运算符变异规则，并通过相关组合构造出多个语义扰动方法．在该研究问题中，我们控制了变异的比例，分别对代码数据的 25%、50%、75% 和 100% 进行变异，以观察不同语义扰动比例下 CodeScore-R 和现有基准指标的鲁棒性表现．在实验中，我们仅对能通过测试用例的代码进行语义扰动．如果扰动成功，我们将对应的 pass@1 值变为 0．为了避免实验结果的偶然性，我们同样选择了五个不同的随机种子，并相应地为每个预测代码生成五种变异代码．然后，我们计算每个变异代码与参考代码的评估指标，并在表 7 至表 10 中分别给出在 Java 代码生成任务、Python 代码生成任务、Python-to-Java 代码迁移任务和 Java-to-Python 代码迁移任务上的 MAE 拟合均值．

Table 7 Comparison Results of CodeScore-R with Baselines for Semantic Perturbation on Java Code Generation Task

表 7 CodeScore-R 指标与基准指标在 Java 代码生成任务上针对语义扰动的对比结果

| 指标 | Java 代码生成 | | | |
|---|---|---|---|---|
| | Top25% | Top50% | Top75% | Top100% |
| BLEU | 0.460 | 0.471 | 0.473 | 0.450 |
| ROUGE-L | 0.441 | 0.476 | 0.499 | 0.498 |
| ChrF | 0.489 | 0.575 | 0.647 | 0.673 |
| ED | 0.491 | 0.572 | 0.644 | 0.675 |
| WeightBLEU | 0.473 | 0.526 | 0.565 | 0.559 |
| SyntaxMatch | 0.481 | 0.529 | 0.565 | 0.570 |
| DataflowMatch | 0.476 | 0.546 | 0.587 | 0.590 |
| CodeBLEU | 0.473 | 0.518 | 0.547 | 0.542 |
| CrystalBLEU | 0.461 | 0.517 | 0.560 | 0.582 |
| CodeBERTScore | 0.515 | 0.660 | 0.788 | 0.872 |
| CodeScore-R | **0.326** | **0.304** | **0.320** | **0.335** |

Table 8 Comparison Results of CodeScore-R with Baselines for Semantic Perturbation on Python Code Generation Task

表 8 CodeScore-R 指标与基准指标在 Python 代码生成任务上针对语义扰动的对比结果

| 指标 | Python 代码生成 | | | |
|---|---|---|---|---|
| | Top25% | Top50% | Top75% | Top100% |
| BLEU | 0.481 | 0.468 | 0.428 | 0.391 |
| ROUGE-L | 0.479 | 0.558 | 0.583 | 0.598 |
| ChrF | 0.484 | 0.550 | 0.575 | 0.592 |
| ED | 0.472 | 0.555 | 0.591 | 0.619 |
| WeightBLEU | 0.489 | 0.508 | 0.489 | 0.471 |
| SyntaxMatch | 0.501 | 0.510 | 0.504 | 0.492 |
| DataflowMatch | 0.482 | 0.576 | 0.611 | 0.637 |
| CodeBLEU | 0.488 | 0.515 | 0.508 | 0.498 |
| CrystalBLEU | 0.484 | 0.552 | 0.577 | 0.596 |
| CodeBERTScore | 0.470 | 0.645 | 0.759 | 0.862 |
| CodeScore-R | **0.301** | **0.317** | **0.320** | **0.312** |

Table 9 Comparison Results of CodeScore-R with Baselines for Semantic Perturbation on Python-to-Java Code Migration Task

表 9 CodeScore-R 指标与基准指标在 Python-to-Java 代码迁移任务上针对语义扰动的对比结果

| 指标 | Python-to-Java 代码迁移 | | | |
|---|---|---|---|---|
| | Top25% | Top50% | Top75% | Top100% |
| BLEU | 0.377 | 0.487 | 0.586 | 0.722 |
| ROUGE-L | 0.342 | 0.451 | 0.555 | 0.681 |
| ChrF | 0.378 | 0.531 | 0.678 | 0.851 |
| ED | 0.378 | 0.545 | 0.711 | 0.901 |
| WeightBLEU | 0.377 | 0.501 | 0.614 | 0.759 |
| SyntaxMatch | 0.395 | 0.499 | 0.606 | 0.738 |
| DataflowMatch | 0.359 | 0.480 | 0.615 | 0.769 |
| CodeBLEU | 0.377 | 0.492 | 0.605 | 0.747 |
| CrystalBLEU | 0.352 | 0.484 | 0.615 | 0.769 |
| CodeBERTScore | 0.387 | 0.567 | 0.747 | 0.947 |
| CodeScore-R | **0.243** | **0.307** | **0.381** | **0.456** |

Table 10 Comparison Results of CodeScore-R with Baselines for Semantic Perturbation on Java -to- Python Code Migration task

表 10 CodeScore-R 指标与基准指标在 Java-to-Python 代码迁移任务上针对语义扰动的对比结果

| 指标 | Java-to-Python 代码迁移 | | | |
|---|---|---|---|---|
| | Top25% | Top50% | Top75% | Top100% |
| BLEU | 0.403 | 0.461 | 0.524 | 0.607 |
| ROUGE-L | 0.360 | 0.523 | 0.696 | 0.873 |
| ChrF | 0.378 | 0.504 | 0.641 | 0.787 |
| ED | 0.365 | 0.525 | 0.695 | 0.867 |
| WeightBLEU | 0.400 | 0.464 | 0.538 | 0.631 |
| SyntaxMatch | 0.449 | 0.475 | 0.518 | 0.581 |
| DataflowMatch | 0.395 | 0.471 | 0.586 | 0.715 |
| CodeBLEU | 0.412 | 0.468 | 0.541 | 0.634 |
| CrystalBLEU | 0.368 | 0.523 | 0.683 | 0.851 |
| CodeBERTScore | 0.355 | 0.552 | 0.757 | 0.961 |
| CodeScore-R | **0.197** | **0.257** | **0.303** | **0.371** |

从表 7 到表 10 的结果中可以观察到，现有指标在面对语义扰动时受到的影响最严重，尤其是 CodeBERTScore 指标，在代码迁移任务上的 MAE 误差值超过 0.94. 一方面是其模型不具备识别代码中微小语义误差的能力，导致计算的相似度过高；另一方面是 CodeBERTScore 计算的是每个词素之间的相似度，导致改变语义的关键词素信息被忽视，以图 6 为例，我们根据 CodeBERTScore 计算代码段 "a = a+b" 和 "a = a-b" 之间的相似度，其计算公式为 (0.995 + 0.986 +0.990 +0.730 + 0.998) / 5 = 0.9398. 其中，关键的语义词素为 "-"，但是在 CodeBERTScore 的计算过程中没有考虑为该词素赋予更大的权重，导致了计算结果的虚高.

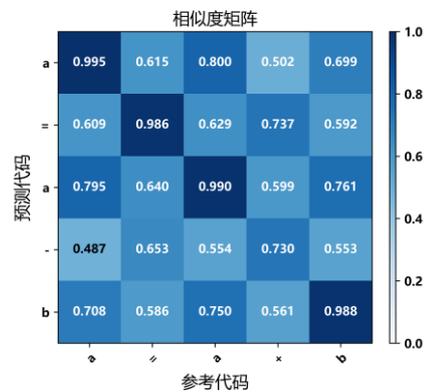

Fig. 6 Process of CodeBERTScore Calculation

图 6 CodeBERTScore 计算过程图

相比于现有指标存在一定的不鲁棒问题，

CodeScore-R 在下游任务上都能表现出对语义扰动具有良好的鲁棒能力. 一方面，这得益于本文提出的 ConCE 框架，让模型具有区分语义差异的能力，另一方面也得益于使用的编译器进行预测代码语法正确的判断，能够避免变异后有语法问题的代码对指标计算的干扰.

### 5.5 RQ5 结果分析

**RQ5**：不同的语义提取方法对 CodeScore-R 有什么影响？

本文通过提取[CLS]的语义向量并经过 ReLU 函数进行激活来提取代码的语义信息. 为证明该提取方法的优势，本文将其与其余三种常见的语义提取方法进行比较：

1）last-avg：该方法对隐藏向量 $H$ 的最后一层进行语义提取，并通过均值化处理得到语义向量.

2）first-last-avg：该方法对隐藏向量 $H$ 的第一层和最后一层进行语义提取，并通过均值化处理得到语义向量.

3）CLS：该方法对隐藏向量 $H$ 最后一层的[CLS]进行语义提取，不额外加入激活函数层.

本文将这三种语义提取方法和本文提出的语义提取方法在 S2S 的设定下进行比较，结果见表 11 和表 12. 从表中可以分析出，在针对 Python 编程语言的下游任务中，本文提出的[CLS]+ReLU 能够取得最优结果，而在 Java 编程语言的下游任务中，first-last-avg 方法可能取得更好的性能结果，因此，可以将语义提取方法当作 CodeScore-R 的超参数，以更灵活地为下游任务进行评估.

## 6 讨论

### 6.1 鲁棒性假设分析

先前的实验结果表明 CodeScore-R 的指标误差均小于现有基准指标，但是并没有直观地分析该指标满足对代码合成评估指标的鲁棒性提出的三个假设. 为进一步探究 CodeScore-R 是否满足对代码合成评估指标的鲁棒性提出的三个假设，本节旨在通过直观的分数值对比来进行验证.

具体而言，分别给出了 CodeScore-R 和选择的基准指标在 Java 代码生成、Python 代码生成、Python-to-Java 代码迁移和 Java-to-Python 代码迁移的计算分数. 其中，"原始"意为不对预测代码进行扰动，"词素扰动"意为对所有的预测代码进行草图化处理，而参考代码保持不变，"语法扰动"意为对所有的预测代码进行等价语法转换，而参考代码保持不变，"语义扰动"意味对所有的预测代码进行变异，而参考代码保持不变. 这些扰动的方法与实验部分的设置保持一致.

从表中可以分析出，无论是在代码生成任务还是代码合成任务上，现有的指标均不能同时满足提出的三种假设. 其中，大部分指标满足假设 2，即当生成的代码中发生语法等价转换时，指标分数应基本不变. 然而，它们对于词素的扰动和语义的扰动并不敏感，以 CodeBERTScore 在 Java-to-Python 代码迁移任务上的指标分数为例，其在"原始"条件下的分数计算为 0.966，但是在"词素扰动"条件下的分数为 0.803，并不符合假设 1；此外，在"语义扰动"条件下，其指标为 0.960，表明该指标不能真实分辨出代码的功能语义，即不符合假设 3.

相比之下，CodeScore-R 能同时满足提出的三种假设. 特别是假设 3，以 Java-to-Python 代码迁移任务上的指标分数为例，CodeScore-R 在"语义扰动"条件下计算出的指标分数为 0.371，远低于其在"原始"条件下计算出的指标分数 0.980，符合上述假设. 因此，CodeScore-R 相较于其他基准指标，具有更强的鲁棒性.

Table 11 Comparison Results of Different Semantic Extraction Methods for CodeScore-R on Code Generation Tasks

表 11 不同的语义提取方法在代码生成任务上对 CodeScore-R 的对比结果

| 方法 | Java 代码生成 | | | | | Python 代码生成 | | | | |
| --- | --- | --- | --- | --- | --- | --- | --- | --- | --- | --- |
| | MAE | Accuracy | Precision | Recall | F1-score | MAE | Accuracy | Precision | Recall | F1-score |
| last-avg | 0.323 | 0.677 | 0.74 | 0.813 | 0.774 | 0.341 | 0.659 | 0.761 | 0.761 | 0.761 |
| first-last-avg | **0.311** | **0.689** | 0.744 | 0.83 | 0.785 | 0.323 | 0.677 | 0.762 | 0.795 | 0.778 |
| CLS | 0.323 | 0.677 | **0.748** | 0.795 | 0.771 | 0.354 | 0.646 | 0.761 | 0.735 | 0.748 |
| CLS+ReLU | 0.317 | 0.683 | 0.717 | **0.884** | **0.792** | **0.287** | **0.713** | **0.765** | **0.863** | **0.811** |

Table 12 Comparison Results of Different Semantic Extraction Methods for CodeScore-R on Code Migration tasks

表 12 不同的语义提取方法在代码迁移任务上对 CodeScore-R 的对比结果

| 方法 | Python-to-Java 代码迁移 | | | | | Java-to-Python 代码迁移 | | | | |
| --- | --- | --- | --- | --- | --- | --- | --- | --- | --- | --- |
| | MAE | Accuracy | Precision | Recall | F1-score | MAE | Accuracy | Precision | Recall | F1-score |
| last-avg | 0.175 | 0.825 | 0.828 | 0.988 | 0.901 | 0.135 | 0.865 | 0.876 | 0.983 | 0.926 |
| first-last-avg | **0.17** | **0.83** | **0.829** | 0.994 | **0.904** | 0.13 | 0.87 | 0.877 | 0.988 | 0.929 |
| CLS | 0.18 | 0.82 | 0.824 | 0.988 | 0.898 | 0.14 | 0.86 | 0.876 | 0.977 | 0.923 |
| CLS+ReLU | 0.175 | 0.825 | 0.821 | **1.000** | 0.902 | **0.125** | **0.875** | **0.878** | **0.994** | **0.932** |

Table 13 Comparison Score Results on Code Generation tasks

表 13 代码生成任务上不同指标分数对比结果

| 指标 | Java 代码生成 | | | | Python 代码生成 | | | |
| --- | --- | --- | --- | --- | --- | --- | --- | --- |
| | 原始 | 词素扰动 | 语法扰动 | 语义扰动 | 原始 | 词素扰动 | 语法扰动 | 语义扰动 |
| BLEU | 0.459 | 0.271 | 0.457 | 0.456 | 0.395 | 0.200 | 0.411 | 0.417 |
| ROUGE-L | 0.655 | 0.385 | 0.478 | 0.504 | 0.618 | 0.550 | 0.416 | 0.626 |
| ChrF | 0.645 | 0.460 | 0.678 | 0.680 | 0.589 | 0.353 | 0.582 | 0.621 |
| ED | 0.663 | 0.581 | 0.673 | 0.681 | 0.624 | 0.521 | 0.641 | 0.649 |
| WeightBLEU | 0.569 | 0.338 | 0.567 | 0.566 | 0.476 | 0.240 | 0.498 | 0.498 |
| SyntaxMatch | 0.639 | 0.639 | 0.620 | 0.580 | 0.513 | 0.497 | 0.502 | 0.507 |
| DataflowMatch | 0.627 | 0.630 | 0.622 | 0.612 | 0.622 | 0.621 | 0.618 | 0.627 |
| CodeBLEU | 0.573 | 0.470 | 0.567 | 0.553 | 0.501 | 0.389 | 0.507 | 0.512 |
| CrystalBLEU | 0.626 | 0.436 | 0.571 | 0.589 | 0.583 | 0.437 | 0.536 | 0.622 |
| CodeBERTScore | 0.879 | 0.778 | 0.875 | 0.878 | 0.881 | 0.781 | 0.892 | 0.897 |
| CodeScore-R | 0.841 | 0.841 | 0.813 | 0.341 | 0.805 | 0.805 | 0.806 | 0.349 |

Table 14 Comparison Score Results on Code Migration tasks

表 14 代码迁移任务上不同指标分数对比结果

| 指标 | Python-to-Java 代码迁移 | | | | Java-to-Python 代码迁移 | | | |
| --- | --- | --- | --- | --- | --- | --- | --- | --- |
| | 原始 | 词素扰动 | 语法扰动 | 语义扰动 | 原始 | 词素扰动 | 语法扰动 | 语义扰动 |
| BLEU | 0.779 | 0.320 | 0.712 | 0.717 | 0.670 | 0.194 | 0.617 | 0.607 |
| ROUGE-L | 0.890 | 0.489 | 0.657 | 0.679 | 0.891 | 0.740 | 0.481 | 0.873 |
| ChrF | 0.862 | 0.479 | 0.841 | 0.850 | 0.795 | 0.356 | 0.725 | 0.786 |
| ED | 0.914 | 0.724 | 0.879 | 0.900 | 0.882 | 0.646 | 0.847 | 0.867 |
| WeightBLEU | 0.815 | 0.343 | 0.750 | 0.754 | 0.695 | 0.207 | 0.653 | 0.631 |
| SyntaxMatch | 0.791 | 0.791 | 0.767 | 0.734 | 0.565 | 0.569 | 0.534 | 0.555 |
| DataflowMatch | 0.883 | 0.850 | 0.875 | 0.864 | 0.694 | 0.670 | 0.682 | 0.693 |
| CodeBLEU | 0.817 | 0.576 | 0.776 | 0.767 | 0.656 | 0.410 | 0.622 | 0.621 |
| CrystalBLEU | 0.867 | 0.507 | 0.753 | 0.768 | 0.853 | 0.555 | 0.658 | 0.851 |
| CodeBERTScore | 0.966 | 0.844 | 0.943 | 0.946 | 0.966 | 0.803 | 0.947 | 0.960 |
| CodeScore-R | 0.980 | 0.980 | 0.945 | 0.450 | 0.980 | 0.980 | 0.968 | 0.371 |

## 6.2 局限性分析

本文提出了一种自动化鲁棒指标 CodeScore-R，用于评估代码合成功能的准确性。然而，该指标仍然存在一些局限性。一方面，CodeScore-R 需要依赖编译器来检测预测代码的语法正确性，而现有的基准指标则不存在上述限制，这可能导致 CodeScore-R 与基准指标的比较存在一定的不公平性。

另一方面，由于深度神经网络本身存在"各向异性"（Anisotropic)的特性，在一些情况下，即使两个代码片段存在一定的语义差异，但由于它们之间仍然具有一些共享特征，导致其在余弦相似度计算时一般会得到较高的分值。

## 7 有效性威胁

### 7.1 内部有效性威胁

第一个内部有效性威胁因素是 CodeScore-R 指标在实现过程中可能存在潜在缺陷，包括代码的草图化处理方法、正负样本的构造和模型的表征学习等方面．为了减轻这些缺陷的影响，我们对处理后的代码进行了检验，以验证其是否能通过测试用例．在模型的表征学习方面，我们采用了 PyTorch 框架[5]和成熟的第三方库 transformers[6]进行实现，以确保实现的正确性和可靠性．

第二个内部有效性威胁是选择的基准指标是否合理．本文选择的基准指标在先前的代码合成领域研究中已经被广泛采用，并且我们根据公开的第三方库进行了统一实验，以确保结果的准确性和可比性．

### 7.2 外部有效性威胁

外部有效性是指研究结果在多大程度上可以推广到实验所使用的数据集之外．在我们的研究中，我们在两个下游任务上进行了实验，并同时关注 Java 和 Python 编程语言．对于 Java 和 Python 编程语言的其他数据集，我们的 CodeScore-R 指标可以直接使用．然而，对于其他编程语言的代码合成任务，需要先在相应编程语言的数据集上进行ConCE表征学习，然后才能应用于该编程语言的代码合成任务中．

此外，在我们的研究中，我们仅调研了 ChatGPT 模型而未考虑传统的方法，这是因为传统方法在代码生成和迁移任务中存在的一些限制和挑战，例如代码的复杂性和语义的多样性，传统的方法一般难以解决这些问题，导致它们的 pass@1 指标不高．因此，我们选择了 ChatGPT 这样的大规模语言模型作为验证工具，因为它具有更强大的语言理解和生成能力，可以应对复杂的代码生成和迁移任务。

### 7.3 构造有效性威胁

构造有效性威胁是指实证研究中使用的评价指标是否真实反映了评估方法的性能．在本文的研究中，考虑了 MAE 作为评判指标对功能准确性的拟合程度．此外，为了避免假阳性或者假阴性数据过多而导致的误差虚低的情况，本文额外将 CodeScore-R 与功能准确性的拟合程度视为分类任务，根据准确率、精度、召回率和 f1-score 来进一步评估 CodeScore-R 指标的性能．

---

[5] https://pytorch.org/
[6] https://github.com/huggingface/transformers

## 8 总结与展望

本文旨在提出一种新的代码质量评估指标 CodeScore-R，该指标基于 UniXcoder 和对比学习．实验结果证明，CodeScore-R 相对于现有基准指标在功能准确性上拟合程度更好．此外，CodeScore-R 与现有基准指标相比，在代码词素扰动、代码语法扰动和代码变异扰动下的鲁棒性表现更好．

在后续研究工作中，我们将在更大规模、更多样性的数据集上验证 CodeScore-R 的性能，同时考虑能力更强大的代码模型作为底座模型，以及进一步探索和改进其在特定领域或特定编程语言上的适用性．

同时，在构造正负样本的方法上也将进行进一步的探索，丰富正负样本的多样性，以适应其在真实数据上的评估．

此外，本文仅在 ChatGPT 模型上进行实证研究，在未来工作中，我们将对传统的方法和基于大规模语言模型的方法进行统一验证，以进一步探索和比较 CodeScore-R 在不同方法中的性能和效果．

**作者贡献声明**：作者一提出了算法思路，负责完成实验并撰写论文，作者二对实验思路提出改进和指导，并给出修改意见，作者三提出指导意见并修改论文，作者四帮助推进实验并检查实验结果．

## 参考文献

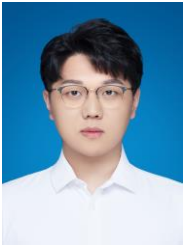

**Yang Guang,** born in 1997. PhD. Student Member of China Computer Federation. His main research interests include code generation, code translation, and robustness in code synthesis.

杨光，1997 年生，博士研究生，CCF 学生会员．主要研究方向为代码生成、代码迁移和代码合成中的鲁棒性研究．

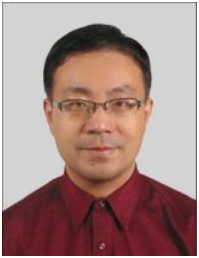

**Zhou Yu,** born in 1981. PhD, professor, PhD supervisor. Senior Member of China Computer Federation. His research interests mainly include software evolution analysis, mining software repositories, software architecture, and reliability analysis. (zhouyu@nuaa.edu.cn)

周宇，1981 年生，博士，教授，博士生导师，CCF 会员．主要研究方向为软件演化分析、挖掘软件库、软件架构和可靠性分析．

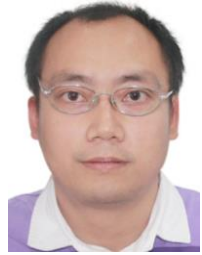

**Chen Xiang,** born in 1980. PhD, associate professor. Member of China Computer Federation. His main research interests include software defect prediction, software defect location, regression testing and combination testing.

陈翔，1980 年生，博士，副教授，CCF 会员．主要研究方向为软件缺陷预测,软件缺陷定位,回归测试和组合测试．

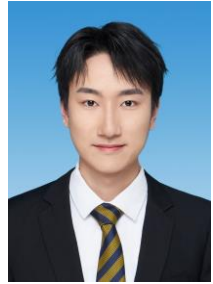

**Zhang Xiangyu**, born in 1999. Postgraduate student. His main research interests include code generation and code pre-trained model.

张翔宇，1999 年生，硕士研究生，主要研究方向为代码生成及代码预训练模型